# Improving statistical learning methods via features selection without replacement sampling and random projection

Sulaiman khan[1], Muhammad Ahmad[1], Fida Ullah[1] Carlos Aguilar Ibañez[1] and José Eduardo Valdez Rodriguez[1*]

[1] Centro de Investigación en Computación, Instituto Politécnico Nacional (CIC-PN),

Mexico City 07738, Mexico; skhan2024@cic.ipn.mx; mahmad2024@cic.ipn.mx (M.A.);

carlosaguilari@cic.ipn.mx

* Correspondence: jvaldezr2018@cic.ipn.mx

**Abstract:** Cancer is fundamentally a genetic disease characterized by genetic and epigenetic alterations that disrupt normal gene expression, leading to uncontrolled cell growth and metastasis. High-dimensional microarray datasets pose challenges for classification models due to the "small *n*, large *p*" problem, resulting in overfitting. This study makes three different key contributions: 1) we propose a machine learning-based approach integrating the Feature Selection Without Replacement (FSWOR) technique and a projection method to improve classification accuracy. 2) We apply the Kendall statistical test to identify the most significant genes from the brain cancer microarray dataset (GSE50161), reducing the feature space from 54,675 to 20,890 genes.3) we apply machine learning models using k-fold cross validation techniques in which our model incorporates ensemble classifiers with LDA projection and Naïve Bayes, achieving a test score of 96%, outperforming existing methods by 9.09%. The results demonstrate the effectiveness of our approach in high-dimensional gene expression analysis, improving classification accuracy while mitigating overfitting. This study contributes to cancer biomarker discovery, offering a robust computational method for analyzing microarray data.

**Keywords:** Brain Cancer Gene expression, Machine learning; SVM, Naïve Bayes, Logistics Regression, DT, KNN, dimension reduction, PCA, LDA, GRP, SRP

## 1. Introduction

Cancer is fundamentally a genetic disease resulting from genetic or epigenetic alterations within somatic cells. These alterations disrupt normal gene expression patterns [3], affecting genes that regulate cell growth, survival, and invasion while suppressing growth-inhibiting genes. The primary mechanism involves the accumulation of mutations, though epigenetic changes such as DNA methylation also play a crucial role. The resulting aberrant gene expression leads to the hallmarks and enabling characteristics of cancer [4]. Most common cancers arise from acquired mutations in somatic cells, while rare hereditary cancer syndromes result from specific germ line mutations.

Cancer-associated genes are categorized into oncogenes (activated, phenotypically dominant) and tumor suppressor genes (inactivated, phenotypically recessive). Oncogene activation can occur through point mutations, gene amplification, or DNA translocation, whereas tumor suppressor genes are inactivated by mutations or promoter silencing [5]. The uncontrolled growth and spread of these abnormal cells define the disease. Cancer is a subset of neoplasms [7], characterized by unregulated cell growth that forms a mass or tumor, potentially spreading diffusely.

Metastasis, the spread of cancer [8], is a multi-step process. It begins with tumor cell detachment from neighboring cells and the surrounding stroma at the primary site. Enzy-matic degradation of the extracellular matrix facilitates the directional movement of individual cells or cell clusters. These cells then invade blood or lymphatic vessels (intravasation) and travel through the circulatory system [9]. Survival during circulation is crucial until they reach a suitable metastatic site, often determined by growth factor availability. At the metastatic site, cells attach to blood vessel endothelium, exit the vessel (extravasation), proliferate, invade, and recruit a new blood supply.

Epithelial-to-mesenchymal transition (EMT) is a key process in the invasion and metastasis of epithelial tumors [10], often followed by a mesenchymal-to-epithelial transition (MET) at the metastatic site [11]. Metastasis patterns to specific organs are not random but are influenced by chemokine receptor expression on tumor cells, guiding them to favorable environments for colony establishment [12]. In short, cancer cells escape their origin and establish new colonies in distant body parts. Distant metastases account for 90% of cancer deaths [13].

The spread can occur through the lymphatic system or bloodstream, leading to the development of new tumors in areas such as the lymph nodes (neck, underarms, groin) or distant organs (liver, bones, brain, lungs)[14]. If cancer spreads from its origin, it is termed metastatic cancer of the original site, not the site of spread [15].

Microarray is a molecular biological method in which tens of thousands of probes demonstrating a given DNA sequence are examined and enumerated to provide a general gene expression profile of multiple biological samples [16]. From a computation viewpoint, single cell RNA sequence studies create a large amount of data with several cells and thousands of genes dimensions [17]. Most single cell RNA sequence data so far belong to small $n$ large $p$ category, where $n$ is the number of sample and $p$ is the number of dimension (Genes) [18]. This violets Gaussian Markov assumption i.e $n > p$. Classification models' effects from overfitting due to large features and a smaller number of samples.

This Study makes the Following Contributions:

- **We apply the Kendall statistical test** to identify the most significant genes from the brain cancer microarray dataset (GSE50161), reducing the feature space from 54,675 to 20,890 genes.
- **We propose a machine learning-based approach** integrating the Feature Selection without Replacement (FSWOR) technique and a projection method to improve classification accuracy.
- **We apply machine learning models using k-fold cross-validation techniques**, where our model incorporates ensemble classifiers with LDA projection and Naïve Bayes, achieving a test score of 96%, outperforming existing methods by 9.09% (Bruno et al.).

## 2. Literature review

Zhang et al. [21] proposed a random projection enhancement (RPE) method to improve surrogate model performance. They applied RPE to least squares support vector regression (LSSVR) and conducted numerical experiments. The results showed improved predictive accuracy, robustness, and optimization performance, even for high-dimensional problems. They further validated RPE's effectiveness on other models and real-world engineering applications.

Hu et al. [22] proposed a hybrid pre-computation-based Heterogeneous Graph Neural Network (RpHGNN) which is used to balances efficiency and low information loss. By using this they introduced a Random Projection Squashing step for linear complexity and a Relation-wise Neighbor Collection component for finer-grained information aggregation. Their experimental results showed state-of-the-art performance on seven benchmark datasets while being 230% faster than the best baseline. Their method outperformed both pre-processing-based and end-to-end models.

Fabiani et al. [23] investigate Best Approximation for Feedforward Neural Networks (FNNs) using Random Projection Neural Networks (RPNNs). They show that RPNNs, with fixed internal weights and non-polynomial activation functions, achieve exponential convergence in function approximation. Five benchmark tests demonstrate their effectiveness, achieving performance comparable to Legendre Polynomials. This highlights RPNNs' potential for efficient and accurate function approximation.

Li et al. [24] proposes a probabilistic framework for sequential random projection, addressing challenges in sequential decision-making under uncertainty. They construct a stopped process to analyze sequential concentration events and derive a non-asymptotic probability bound. This extends the Johnson-Lindenstrauss lemma to a martingale setting, contributing to random projection and sequential analysis.

Asi et al. [25] proposes Projection Unit, a framework for efficient, private mean estimation by projecting inputs to random low-dimensional subspaces. This method achieves near-optimal error with reduced communication and computational costs. Experiments show it performs similarly to optimal algorithms in private mean estimation and federated learning.

McDonnell et al. [26] proposes a novel approach for continual learning (CL) with pre-trained models, addressing catastrophic forgetting by using frozen Random Projection layers and class-prototype accumulation. This method enhances linear separability and decorrelates class-prototypes to reduce distribution gaps. Experiments show that their approach reduces error rates by 20%-62% on class-incremental datasets without using rehearsal memory.

Kumaran et al. [27] explores the use of local random quantum circuits for dimensionality reduction of large low-rank datasets, leveraging the random projection method. They show that quantum circuits with short depths perform comparably to classical principal component analysis on image datasets like MNIST and CIFAR-100. Benchmarking quantum random projection against classical methods, they demonstrate its effectiveness in reducing dimensions and computing von Neumann entropies, as well as implementing singular value decomposition for large matrices.

**Table 1.** Prior studies on Brain cancer detection.

| Reference | Supervised Methods | Techniques | C.V score |
| --- | --- | --- | --- |
| Feltes et al [1] | **SVM**, NB, RF, DT, MLP | **PCA** | 0.88 |
| proposed | DT, **NB**, SVM, LR, KNN | FSWOR(PCA), **FSWOR(LDA)**, FSWOR(GRP), FSWOR(SRP) | 0.96 |

## 3. Methodology

*3.1. Brain Cancer Dataset*

We sourced our dataset from Kaggle (https://www.kaggle.com/datasets/bruno-grisci/brain-cancer-gene-expression-cumida), which contains gene expression data for brain cancer. The dataset comprises a total of 130 samples, with 13 classes as normal and 117 classes as abnormal tissues. In the abnormal cases, there are four different cancer types such as **Ependymoma, Glioblastoma, Medulloblastoma, and Pilocytic Astrocytoma**. The corpus is in well-structured format which contains **54,675 independent gene expression (features)**. This extensive data allows for in-depth analysis of different brain cancer types.

We apply Kendall statistical test on this dataset. We obtained 20,890 significant genes. We apply normalization scaling technique after Kendall statistical test.

*3.2. Randomly Features selection*

The figure 1 outlines a structured methodology for developing machine learning models to analyze brain cancer data. The workflow begins with pre-processing, where raw data is refined through techniques such as normalization and handling of missing values to ensure robustness. Feature selection is rigorously addressed via the Kendall Statistical Test, which identifies statistically significant biomarkers, and random feature selection (without replacement) to enhance model diversity in each ensemble [20] and reduce dimensionality.

Subsequently, projection techniques are applied to distill high-dimensional data into interpretable subspaces. These include:
- Principal Component Analysis (PCA) for variance maximization,
- Linear Discriminant Analysis (LDA) to optimize class separability,
- Gaussian Random Projection (GRP) and Sparse Random Projection (SRP) to preserve pairwise distances while improving computational efficiency.

The machine learning models such as Logistic Regression (LR), Support Vector Machine (SVM), Decision Tree (DT), Naïve Bays (NB) and K Nearest Neighbour (KNN) are trained [28,29], and validated using a 3-fold cross-validation framework, ensuring rigorous assessment of generalizability as shown in figure 1. Ensemble methods are employed, where individual learners are trained on distinct feature subsets to mitigate overfitting and enhance predictive stability.

Performance metrics—precision, recall, F1-score, and accuracy—are systematically evaluated to quantify diagnostic efficacy. Precision reflects the model's ability to minimize false positives, while recall captures sensitivity in detecting true cancer cases. The F1-score harmonizes these metrics, and accuracy provides an aggregate measure of correctness.

This pipeline integrates statistical rigor, algorithmic diversity, and iterative validation, aiming to advance computational tools for brain cancer diagnosis. By bridging data-driven insights with clinical relevance, the framework underscores the potential of machine learning to augment oncological decision-making, ultimately contributing to precision medicine initiatives.

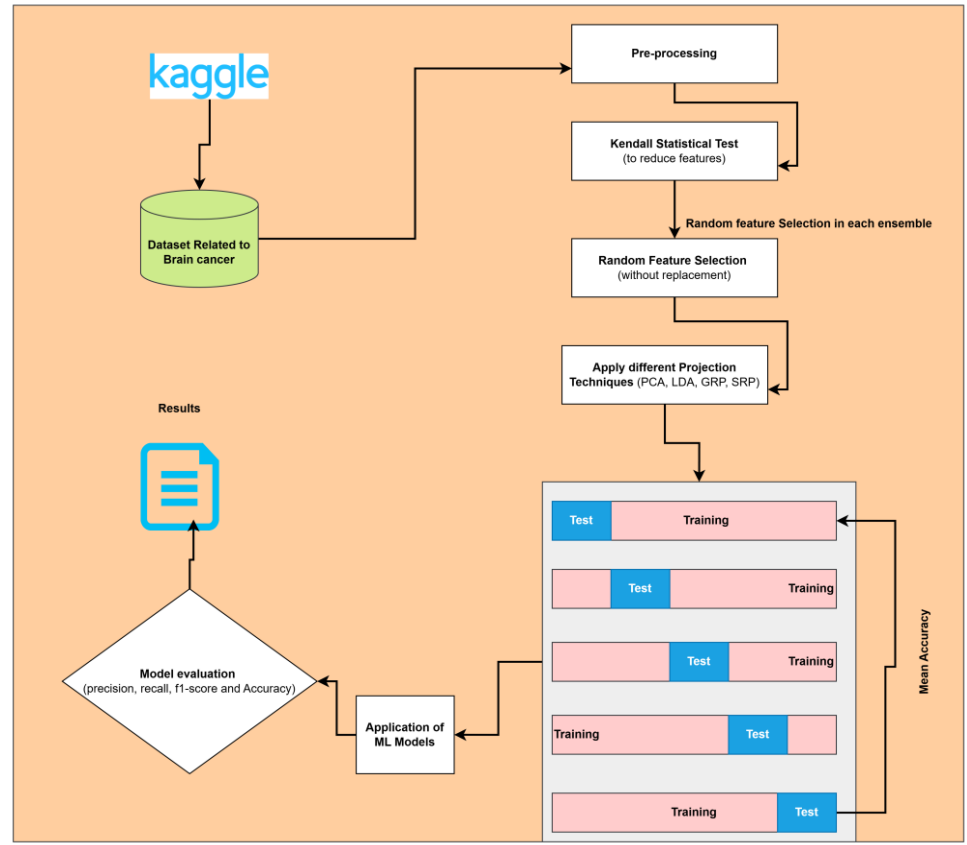

**Fig 1.** Propose architecture and design.

## 4. Results and Analysis

*4.1 Statistical Analysis*

We have mention class distribution, scatterplot before Kendall statistical test and after Kendall statistical test, obtained the most significant features i.e 20890. These features are used in Machine learning models with total samples 130.

Figure 2 describes different classes of Brain cancer type. Ependymoma, Glioblastoma, Medulloblastoma, Normal and Pilocytic_astrocytoma are 46, 34, 22, 13 and 15 classes respectively. 0, 1, 2, 3 and 4 denote Ependymoma, Glioblastoma, Medulloblastoma, Normal and Pilocytic_astrocytoma classes respectively. But Normal class is healthy tissue.

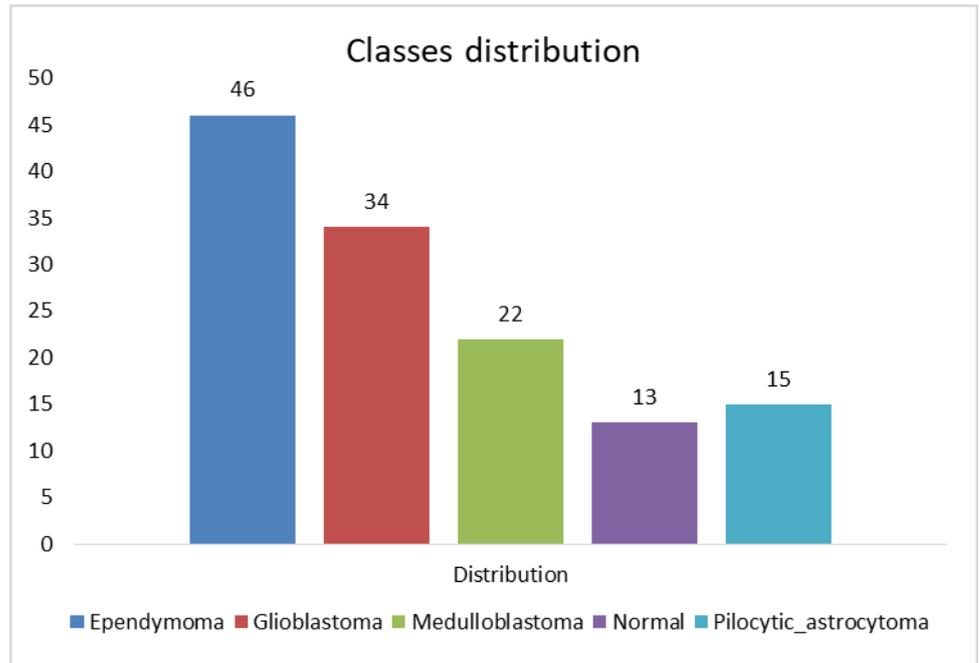

**Fig 2.** Class distribution

Figure 3 visualizes the relationship between two variables, 1007_s_at and 1053_at categorized into multiple classes (labeled 0 to 4). The plot helps in understanding how these classes (Type) are not linearly separable in different groups in the data before Kendall statistical test. In our dataset classes were not linearly separable in different assemblies, to handle this task we employed Kendall statistical test and reduce the feature size to make classes linearly separable.

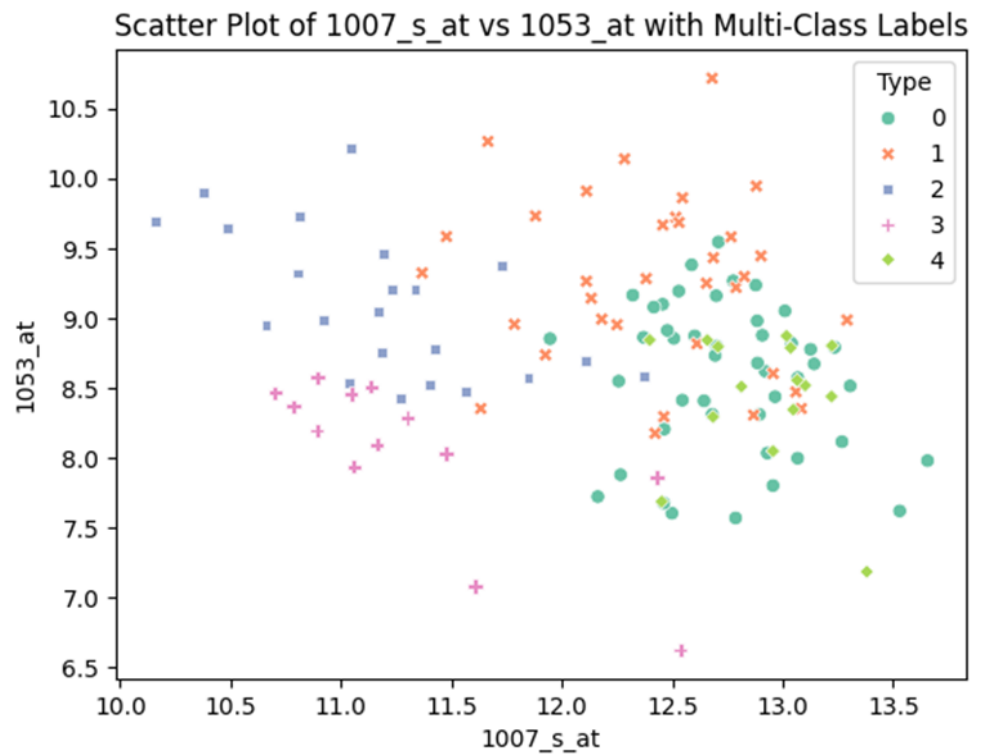

**Fig 3.** Scatter plot before Kendall Statistical Test

4.1.2 Kendall Statistical Test

Figure 4 represent the results of a Kendall Statistical Test, which is a non-parametric test used to assess the association between two measured quantities. The y-axis likely represents the p-values obtained from the test, with values ranging from 5.00E-22 (a very small number indicating a highly significant result) down to 0.00E+00. The x-axis lists different identifiers, possibly representing genes or probes, such as "230763_at," "239515_at," and others.

The p-values are extremely low, suggesting that the associations being tested are highly statistically significant. For example, the p-value of 1.06E-22 is astronomically small, indicating that the likelihood of the observed association occurring by chance is virtually zero. The identifiers on the x-axis correspond to specific data points or samples being analyzed, and the p-values associated with each suggest that there is a strong, statistically significant relationship being measured.

In simpler terms, this figure shows us that the data being analyzed shows very strong evidence of a relationship, and the results are not due to random chance. This could be crucial in fields like genetics or bioinformatics, where identifying significant associations can lead to important discoveries about gene functions or disease mechanisms.

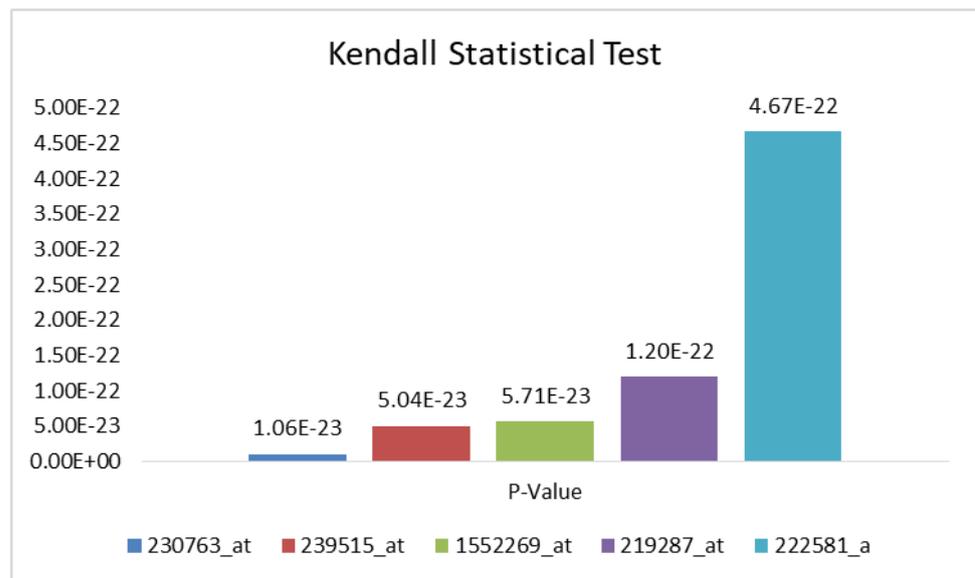

**Fig 4.** Kendall Statistical Test

Figure 5 visualizes the relationship between two variables, 230763_at and 239515_at categorized into multiple classes (labeled 0 to 4). The plot helps in understanding how these classes (Type) interact across features, which can be useful for identifying patterns or clusters in the data after Kendall statistical test.

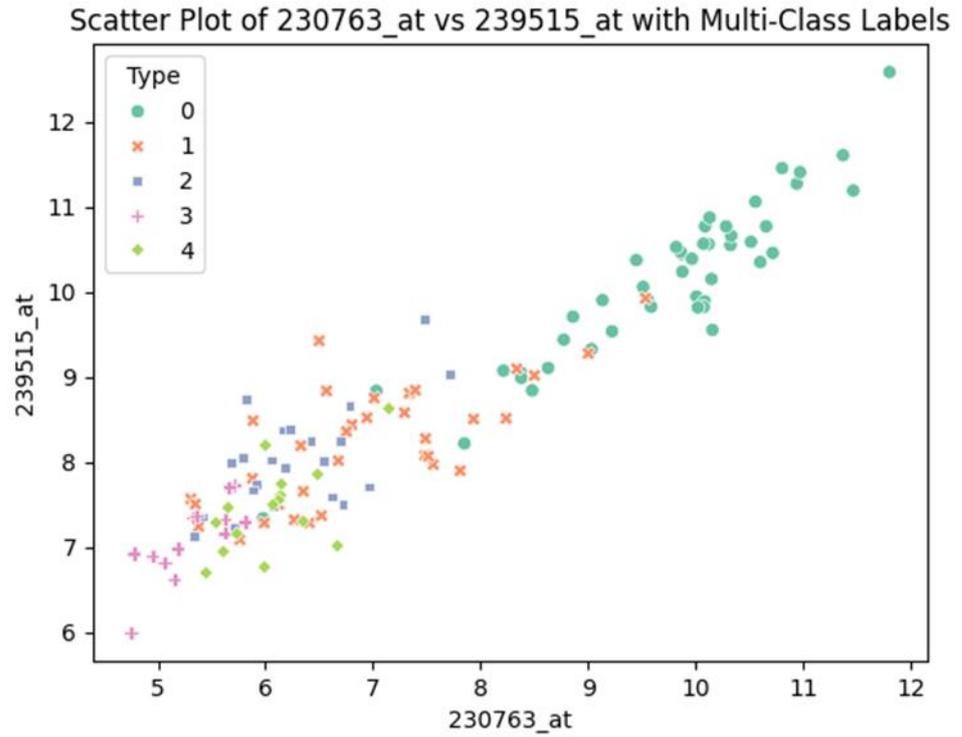

**Fig 5.** Scatter plot after Kendall Statistical Test

Figure 6 demonstrates the histogram of genes i.e. 230763_at, 239515_at and 1552269_at. The 230763_at expression level is very high between 5 and 7, moderate between 9 and 11. The 239515_at expression level is peak between 7 and 9. The 1552269_at expression level is very low between 6 and 8, between 9 and 10. The specific and peak range indicate which genes are more active (higher expression) or deactivated (lower expression) across samples.

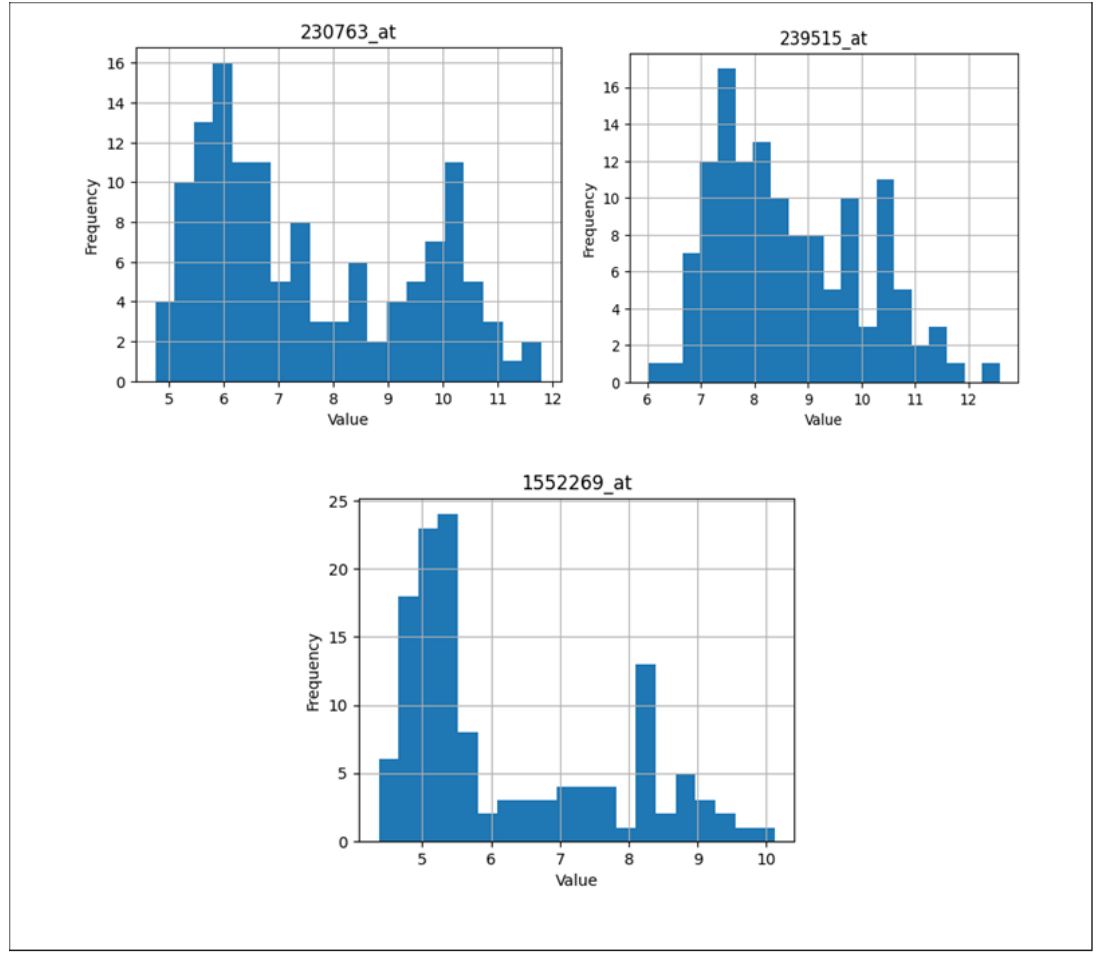

**Fig 6.** Histogram of Genes

*4.2 Machine learning analysis*

4.2.1 Classifier performance

Table 2 presents the performance metrics of five models Decision Tree (DT), Support Vector Machine (SVM), Logistic Regression (LR), K-Nearest Neighbors (KNN), and Naive Bayes (NB) after applying Principal Component Analysis (PCA) for dimensionality reduction. Among the models, SVM stands out with the highest average precision (0.900), recall (0.884), F1-score (0.887), and test accuracy (0.915), showcasing its superior ability to classify data accurately. KNN and Naive Bayes also perform well, achieving balanced metrics and strong accuracy scores (0.896 and 0.892, respectively). Logistic Regression and Decision Tree, while slightly lower in performance, still demonstrate reliable outcomes with precision, recall, F1-scores, and test accuracy values that reflect solid classification capabilities. Overall, the table highlights SVM as the most effective model, followed by KNN and NB, in this PCA-optimized classification task.

**Table 2.** Machine learning Results with PCA

| Models | Precision | Recall | F1-score | Accuracy |
|--------|-----------|--------|----------|----------|
| DT     | 0.837     | 0.834  | 0.827    | 0.862    |
| SVM    | 0.900     | 0.884  | 0.887    | 0.915    |
| LR | 0.856 | 0.838 | 0.840 | 0.885 |
| KNN | 0.873 | 0.873 | 0.871 | 0.896 |
| NB | 0.868 | 0.859 | 0.861 | 0.892 |

Table 3 summarizes the performance of five classification models Decision Tree (DT), Support Vector Machine (SVM), Logistic Regression (LR), K-Nearest Neighbors (KNN), and Naive Bayes (NB) after applying Linear Discriminant Analysis (LDA) for feature extraction. Naive Bayes (NB) achieves the best results overall, with the highest average precision (0.938), recall (0.916), F1-score (0.923), and accuracy (0.938). KNN also performs exceptionally well, closely following NB in precision (0.934), recall (0.900), F1-score (0.905), and accuracy (0.927). SVM and Logistic Regression show robust performance as well, with SVM attaining an accuracy of 0.919, slightly outshining LR. Decision Tree, while effective, lags behind the other models, particularly in F1-score (0.845) and accuracy (0.873). Overall, the results emphasize that NB and KNN benefit from the most LDA, offering the most consistent and accurate classifications.

**Table 3.** machine learning results with LDA

| Models | Precision | Recall | F1-score | Accuracy |
|---|---|---|---|---|
| DT | 0.862 | 0.857 | 0.845 | 0.873 |
| SVM | 0.923 | 0.898 | 0.901 | 0.919 |
| LR | 0.910 | 0.877 | 0.881 | 0.896 |
| KNN | 0.934 | 0.900 | 0.905 | 0.927 |
| NB | 0.938 | 0.916 | 0.923 | 0.938 |

Table 4 highlights the performance metrics of five models Decision Tree (DT), Support Vector Machine (SVM), Logistic Regression (LR), K-Nearest Neighbors (KNN), and Naive Bayes (NB) after using Gaussian Random Projection (GRP) for features transformation. SVM emerges as the top performer, achieving the highest average precision (0.893), recall (0.844), F1-score (0.853), and accuracy (0.873), making it the most reliable model in this context. Logistic Regression and Naive Bayes follow with balanced metrics, delivering accuracy scores of 0.838 and 0.815, respectively. KNN shows moderate performance, with a precision of 0.789 and an accuracy of 0.800. In contrast, the Decision Tree performs poorly compared to the other models, with the lowest precision (0.636), recall (0.600), F1-score (0.592), and accuracy (0.658). These results suggest that SVM is the most effective model under GRP transformation, while DT struggles to adapt effectively.

**Table 4.** machine learning results with GRP.

| Models | Precision | Recall | F1-score | Accuracy |
|---|---|---|---|---|
| DT | 0.636 | 0.600 | 0.592 | 0.658 |
| SVM | 0.893 | 0.844 | 0.853 | 0.873 |
| LR | 0.821 | 0.813 | 0.806 | 0.838 |
| KNN | 0.789 | 0.788 | 0.778 | 0.800 |
| NB | 0.809 | 0.776 | 0.772 | 0.815 |

Table 5 presents the performance of five models Decision Tree (DT), Support Vector Machine (SVM), Logistic Regression (LR), K-Nearest Neighbors (KNN), and Naive Bayes (NB) after applying Sparse Random Projection (SRP) for feature transformation. SVM achieves the highest average performance across all metrics, with precision (0.915), recall

(0.855), F1-score (0.866), and accuracy (0.885), making it the standout model in this setup. Naive Bayes (NB) and Logistic Regression (LR) deliver similar and balanced results, with NB slightly leading in accuracy (0.835) and LR showing a consistent performance across all metrics. KNN demonstrates moderate results, maintaining acceptable scores but falling behind the top-performing models. Decision Tree (DT), however, exhibits the weakest performance, with the lowest precision (0.576), recall (0.567), F1-score (0.559), and accuracy (0.619). Overall, the table underscores SVM's robustness and highlights DT's difficulty in adapting to the SRP transformation.

**Table 5.** machine learning results with SRP.

| Models | Precision | Recall | F1-score | Accuracy |
|---|---|---|---|---|
| DT | 0.576 | 0.567 | 0.559 | 0.619 |
| SVM | 0.915 | 0.855 | 0.866 | 0.885 |
| LR | 0.831 | 0.804 | 0.799 | 0.827 |
| KNN | 0.799 | 0.773 | 0.766 | 0.792 |
| NB | 0.849 | 0.802 | 0.807 | 0.835 |

Table 6 highlights the top-performing models for each feature transformation technique: PCA, LDA, GRP, and SRP. Across these methods, Support Vector Machine (SVM) consistently demonstrates superior performance, achieving the average highest scores in three techniques. Under PCA, SVM achieves a precision of 0.900, recall of 0.884, F1-score of 0.887, and accuracy of 0.915. For LDA, Naive Bayes (NB) emerges as the leader with outstanding results, including a precision of 0.938, recall of 0.916, F1-score of 0.923, and accuracy of 0.938. In both GRP and SRP, SVM once again excels, showcasing its adaptability and reliability across different transformations. This table emphasizes SVM's robustness and versatility, while also acknowledging NB's strong performance under LDA.

**Table 6.** Top performing models in each Projection

| Models | Precision | Recall | F1-score | Accuracy |
|---|---|---|---|---|
| PCA | | | | |
| SVM | 0.900 | 0.884 | 0.887 | 0.915 |
| LDA | | | | |
| NB | 0.938 | 0.916 | 0.923 | 0.938 |
| GRP | | | | |
| SVM | 0.893 | 0.844 | 0.853 | 0.873 |
| SRP | | | | |
| SVM | 0.915 | 0.855 | 0.866 | 0.885 |

The Figure 7 illustrates the testing performance of various models Logistic Regression (LR), Support Vector Machine (SVM), K-Nearest Neighbors (KNN), Naive Bayes (NB), and Decision Tree (DT) using different projection techniques: PCA, LDA, GRP, and SRP.

Naive Bayes (NB) with LDA achieves the highest score (0.94), highlighting its effectiveness with this technique. SVM also performs consistently well, particularly with LDA and PCA, achieving top scores of 0.92 for each. KNN and Logistic Regression deliver stable results across the projection methods, with scores hovering near 0.90 under PCA and LDA but slightly lower with GRP and SRP. Decision Tree lags behind significantly, with its highest score being 0.87 (under PCA and LDA) and its weakest performance under GRP (0.66) and SRP (0.62).

Overall, the figure emphasizes the strength of SVM and NB across different projections and the notable gap in performance for DT under GRP and SRP.

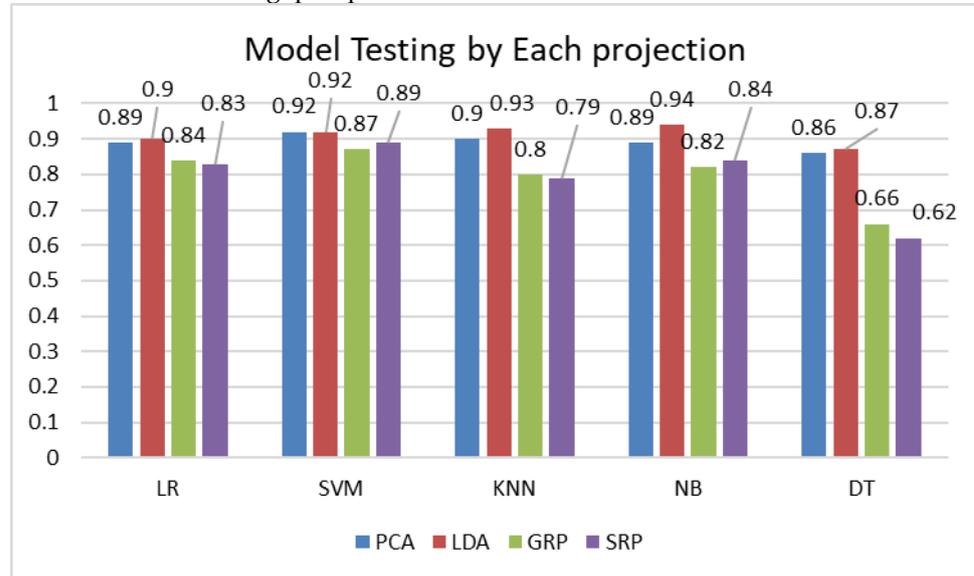

**Fig 7.** Comparisons of different machine learning models using different projection techniques

Figure 8 illustrates the accuracy of a SVM model but using Principal Component Analysis (PCA) for dimensionality reduction. The accuracy trends suggest that PCA might be more stable or effective compared to Sparse Random Projection, as indicated by the higher accuracy values.

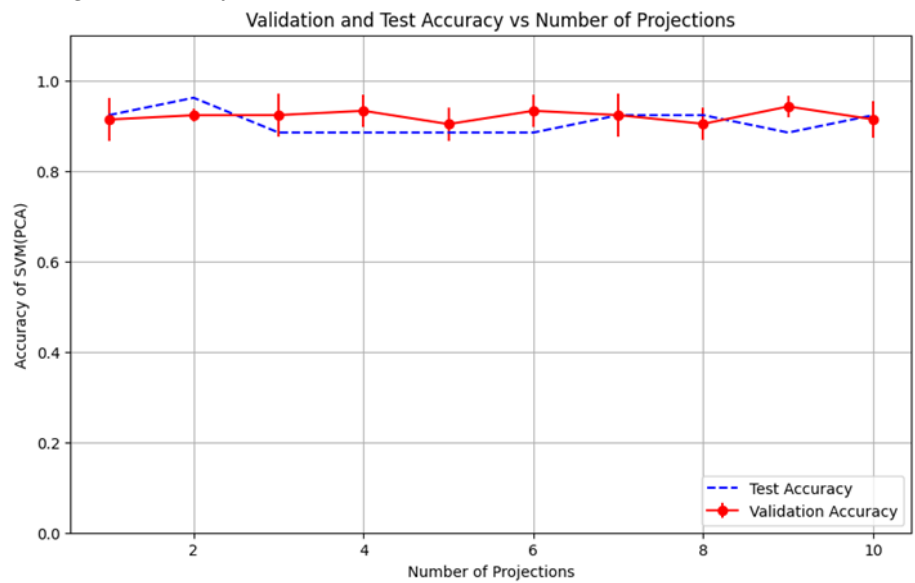

**Fig 8.** Validation and Test accuracy comparison vs number of projections.

Figure 9 represents the accuracy of a Naive Bayes model using Linear Discriminant Analysis (LDA) for dimensionality reduction. The accuracy trends here show how the model performs as the number of projections changes, providing insights into the effectiveness of LDA in this context.

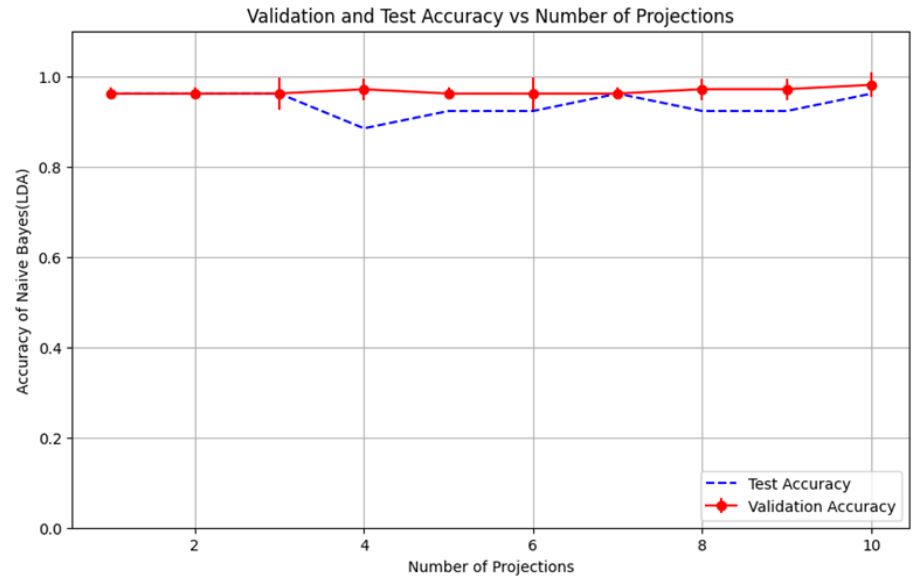

**Fig 9.** Validation and Test accuracy comparison vs number of projections.

Figure 10 depicts the accuracy of an SVM model using Gaussian Random Projection. The graph helps in understanding how this particular projection method affects the model's accuracy across different numbers of projections.

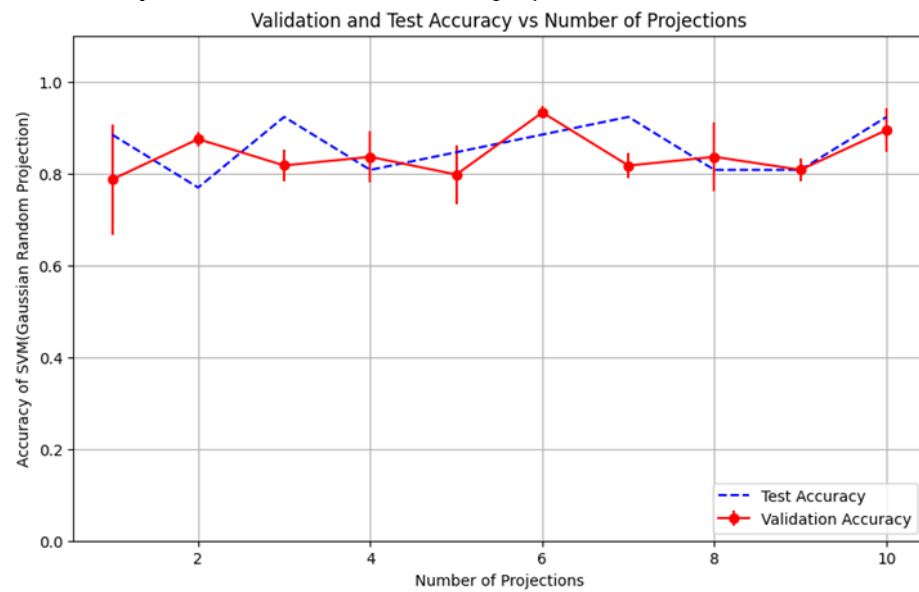

**Fig 10.** Validation and Test accuracy comparison vs number of projections.

Figure 11 shows the accuracy of a Support Vector Machine (SVM) model using Sparse Random Projection for dimensionality reduction. The accuracy is plotted against the number of projections. It appears that as the number of projections increases, the accuracy fluctuates, indicating that the optimal number of projections is crucial for achieving the best model performance.

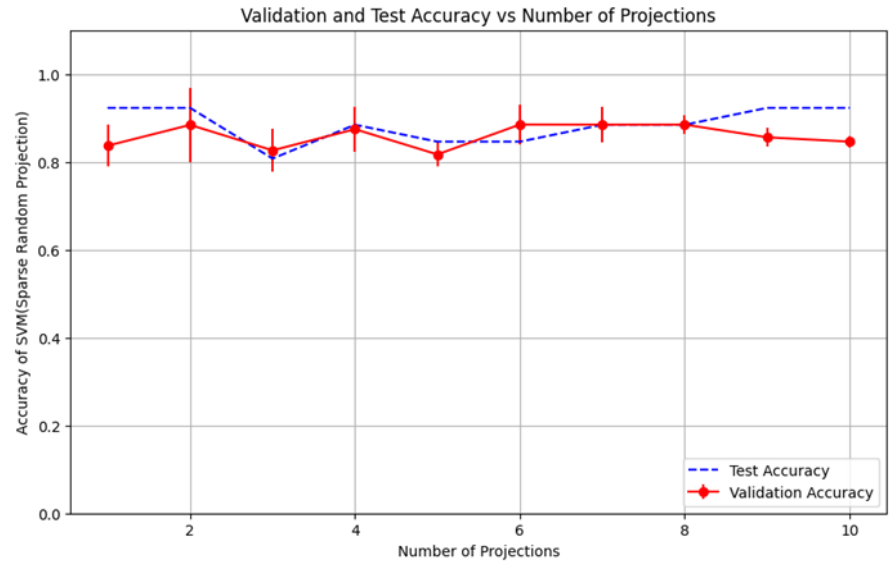

**Fig 11.** Validation and Test accuracy comparison vs number of projections.

Over all these figures collectively highlight the importance of choosing the right dimensionality reduction technique and the optimal number of projections to maximize model accuracy. Each method (SRP, PCA, LDA, GRP) has its own impact on the model's performance, and these visualizations help in making informed decisions about which technique to use based on the specific requirements of the task at hand.

Figure 12 encloses four mean confusion matrices. The mean confusion matrices summarize the classification results by comparing actual vs predicted labels. **Top-left**, SVM (PCA) predicted hundred percent for class 0. Similarly, class 2 and class 3. It predicted 590% correctly classified for class 1 but 220% misclassified against actual class 4. This technique predicted 80% correctly classified for class 4 but 110% misclassified against actual class1. **Top-right**, Naïve Bayes (LDA) predicts 900% correctly for class 0 but 10%, 10% misclassified for class 0 against actual class 1 and class 4 respectively. Similarly, class 2 and class 3 correctly classified. Class 1 has 660% correct predictions but it predicts hundred percent misclassifications against actual class 4. It predicts 190% correctly for class 4 but 30% predictions misclassified for class 4 against actual class 1. **Bottom-left**, SVM (GRP) predicts correctly for class 0 (870%), class 1(490%), class 2 (370%) and class 3 (300%), class 4(80%) but class 0 is misclassified 50%, 10% against actual class 1 and class 4 respectively. Class 2 and class 3 are misclassified 50%, 10% against actual class 1 and class 2 respectively. Class 1 is misclassified 140%, 20% and 30% against actual class 4, class 2 and class 0 respectively. It predicts (150%) correctly for class 4 but 110% misclassified against actual class 1. **Bottom-right**, SVM (SRP) has 880% correct predictions for class 0. Similarly, class 1, class 2, class 3 and class 4 have 460%, 390%,300% and 150%correct prediction respectively but 50%, 120% misclassified for class 0 against actual class 4, class 1. Similarly, class 1 has misclassified prediction 100%, 20% against actual class 4, class 0 respectively. 20%, 10%, 100% misclassified for class 2, class 3 and class 4 against actual class 1, class 2 and class 1 respectively.

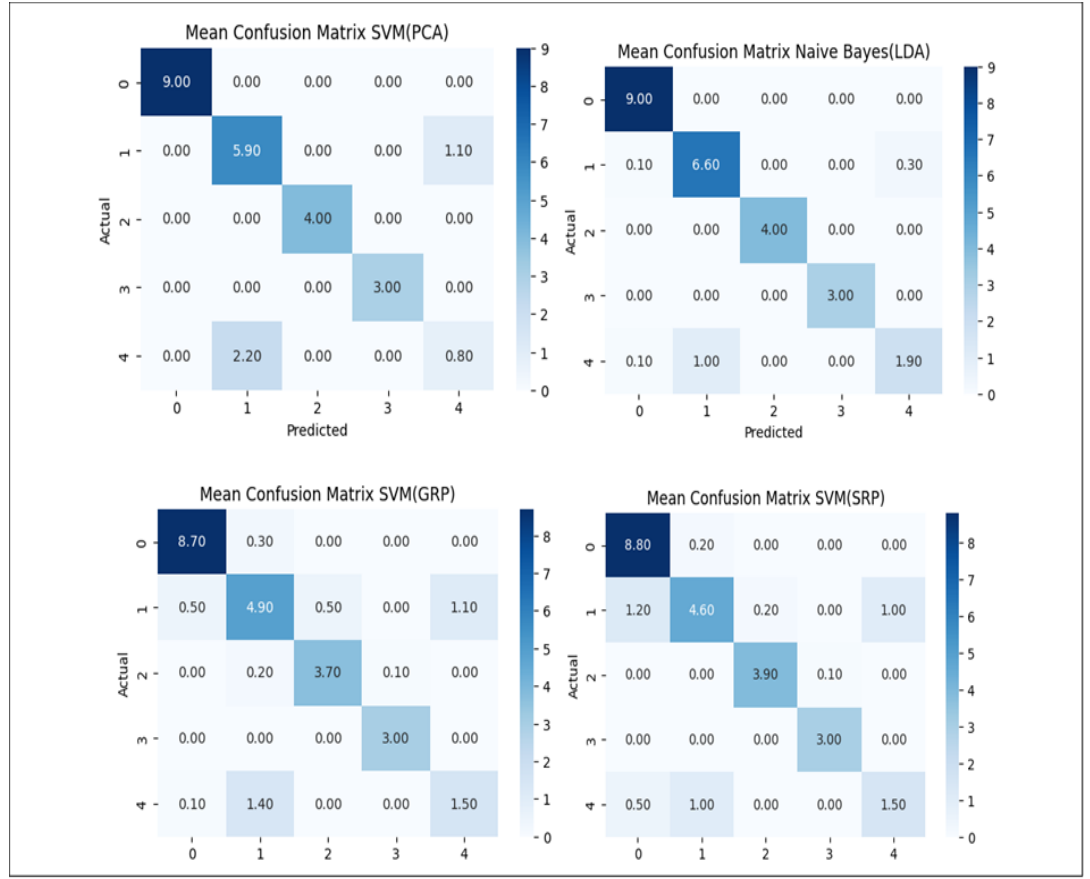

**Fig 12.** Confusion matrix.

## 5. Conclusion

In this study, we proposed a machine learning-based model utilizing the FSWOR technique and a projection method to address overfitting in high-dimensional brain cancer microarray data (GSE50161). We applied the Kendall statistical test to identify 20,890 significant genes from total genes 54,675 across 130 samples. Our approach integrated ensemble classifiers with feature selection without replacement and projection techniques to improve classification accuracy. The proposed model, leveraging LDA projection with Naïve Bayes, outperformed existing methods, achieving a cross-validation score of 96%. This result demonstrates the effectiveness of our approach in high-dimensional gene expression analysis. The ability to extract biologically relevant features enhances interpretability in cancer classification. Our findings highlight the importance of statistical feature selection and dimensionality reduction in microarray data analysis. The proposed method significantly mitigates overfitting while maintaining high predictive performance. Future work may explore deep learning techniques and other projection methods for further improvements. Overall, our study contributes to advancing computational techniques for cancer biomarker discovery and classification.

**Author Contributions:** Conceptualization, S.K., M.A., and J. E.; methodology, M.A., S.K.; software, F.U., S.K ,J. E; validation, J.E. C.A.; formal analysis, F.U., S.K and C.A.; investigation, J.E.; resources, S. K. and M.A.; data curation, S.K., M.A; writing—original draft preparation, M.A. and S.K.; writing—review and editing, M.A. S.K and F.U.; visualization, J.E. and C.A.; supervision, J.E.; project administration, C.A. All authors have read and agreed to the published version of the manuscript.

**Funding:** This research received no external funding.

**Data Availability Statement:** The dataset utilized in this study is publicly available on Kaggle. Kaggle (https://www.kaggle.com/datasets/brunogrisci/brain-cancer-gene-expression-cumida).

**Acknowledgments:** This work was performed with partial support from the Mexican Government through the grant A1-S-47854 of CONACYT, Mexico, and grants 20241816, 20241819, and 20240951 of the Secretaría de Investigación y Posgrado of the Instituto Politécnico Nacional, Mexico. The authors thank CONACYT for the computing resources brought to them through the Plataforma de Aprendizaje Profundo para Tecnologías del Lenguaje of the Laboratorio de Supercómputo of the INAOE, Mexico and acknowledge support of Microsoft through the Microsoft Latin America PhD.

**Conflicts of Interest:** The authors declare no conflicts of interest.